\documentstyle[aps,epsf,rotate]{revtex}
\begin{document}
\draft
\title{Lyapunov instability for a periodic Lorentz gas\\ 
thermostated by deterministic scattering}
\author{K. Rateitschak\footnote{e-mail: katja.rateitschak@molgen.mpg.de}}
\address{Center for Nonlinear Phenomena and Complex Systems, 
Universit\'{e} Libre de Bruxelles,\\ 
Campus Plaine CP 231, Blvd du Triomphe, B-1050 Brussels,
Belgium}
\author{R. Klages\footnote{e-mail: rklages@mpipks-dresden.mpg.de}}
\address{Max Planck Institute for Physics of Complex Systems,\\
N\"othnitzer Str.~38, D-01187 Dresden, Germany}
\date{\today}
\maketitle
\begin{abstract} 
In recent work a deterministic and time-reversible boundary thermostat called
thermostating by deterministic scattering has been introduced for the periodic
Lorentz gas [Phys.\ Rev.\ Lett.\ {\bf 84}, 4268 (2000)]. Here we assess the
nonlinear properties of this new dynamical system by numerically calculating
its Lyapunov exponents. Based on a revised method for computing Lyapunov
exponents, which employs periodic orthonormalization with a constraint, we
present results for the Lyapunov exponents and related quantities in
equilibrium and nonequilibrium. Finally, we check whether we obtain the same
relations between quantities characterizing the microscopic chaotic dynamics
and quantities characterizing macroscopic transport as obtained for
conventional deterministic and time-reversible bulk thermostats.
\end{abstract}


\section{Introduction}
The investigation of nonequilibrium transport processes in many-particle
systems generally requires to model the interaction between a particle and a
thermal reservoir.  A common approach for such a modeling are deterministic
and time-reversible thermostats \cite{EvMo90,Hoov91,MoDe98,KRN98}.
Conventional types of them, such as the Gaussian and the Nos\'e-Hoover
thermostat, are based on introducing a momentum dependent friction coefficient
into the microscopic equations of motion
\cite{HLM82,Ev83,EH83,Nose84b,Hoov85}.
Though the microscopic equations of motion of these systems are
 time-reversible the macroscopic dynamics is irreversible in nonequilibrium
leading to momentum and energy fluxes with well-defined transport coefficients
\cite{HLM82,Ev83,EvHo85,MH87,HHP87,Hoov88,HoB99}, which appears to be a
paradox.  However, investigations of the microscopic dynamics with methods
from dynamical system theory could resolve this paradox by showing that the
microscopic dynamics is nonlinear and highly unstable
\cite{HHP87,HoB99} and leads to a phase space volume contraction 
onto a fractal attractor \cite{MH87,HoMo89,Morr89}. From the analysis of
conventional thermostats, further relations between quantities characterizing
the microscopic dynamics and quantities characterizing macroscopic transport
could be established. At the heart of such relations there is an identity
between phase space volume contraction and thermodynamic entropy
production. On the basis of this identity the Lyapunov exponents could be
related to the transport coefficients of a system, which has been formulated
as the Lyapunov sum rule \cite{MH87,PoHo88,Ch1,Ch2,ECM,DeGP95}.

These characteristic features of thermostated many-particle systems have been
recovered for specific one-particle systems, the Gaussian thermostated
periodic Lorentz gas
\cite{MH87,HoMo89,Ch1,Ch2,DeGP95,LRM94,LNRM95,DettmannMorriss96} and the
Nos\'e-Hoover thermostated periodic Lorentz gas \cite{RKH00}.  The periodic
Lorentz gas consists of a particle that moves through a triangular lattice of
hard disks and is elastically reflected at each disk collision.  It serves as
a standard model in the field of chaos and transport \cite{HoB99,De00}.  The
advantage of a one-particle system is that it reflects more strongly and
transparently the nonequilibrium properties induced by a thermostat.  For this
reason the Lorentz gas appears as an appropriate tool to compare the
properties of nonequilibrium steady states obtained from different
deterministic and time-reversible thermostating mechanisms.  The study of
different models describing the interaction between particles and thermal
reservoir and the identification of their common properties is crucial to
obtain a general characterization of nonequilibrium steady states.

To investigate whether the nonequilibrium properties of conventional
deterministic and time-reversible thermostats are of general validity, or just
characterize these specific types of systems, an alternative deterministic and
time-reversible thermostat called thermostating by deterministic scattering
has been introduced for the periodic Lorentz gas \cite{KRN98,RKN99}.  This
thermostat is based on specifically modeling the energy transfer related to a
microscopic collision process between particle and disk, where the disk mimics
a thermal reservoir with infinitely many degrees of freedom.  In
nonequilibrium under an external electric field this mechanism leads to an on
average constant kinetic energy of the particle resulting in a nonequilibrium
steady state.  Furthermore, the phase space volume contracts onto an attractor
similar to the multifractal attractor found for the Gaussian thermostated
Lorentz gas. However, differences appear in the bifurcation diagram and in the
field dependence of the conductivity.  This alternative thermostat has later
been applied to a heat and shear flow \cite{WKN98}.
 
In this work we focus on the microscopic properties of thermostating by
deterministic scattering in the periodic Lorentz gas by numerically
calculating the Lyapunov exponents.  As quantities from dynamical systems
theory, Lyapunov exponents allow a detailed characterization of the
microscopic stability.  In particular, they will enable us to check the
general validity of relations between quantities from dynamical systems theory
and statistical mechanics as obtained for conventional deterministic and
time-reversible thermostats.  We first explain the algorithm to calculate the
Lyapunov exponents for the Lorentz gas as thermostated by deterministic
scattering. Numerical computations will then show that the standard
Gram-Schmidt orthonormalization has to be modified resulting in a variant of
this method called constraint orthonormalization.  Beside the results for the
Lyapunov exponents we present results for the Kaplan-Yorke dimension and for
the phase space volume contraction.  We compare these results as obtained for
our model with the results as known for the Gaussian thermostated Lorentz gas
\cite{DeGP95,DePH96}, and with results for a heat and shear flow thermostated by
deterministic scattering \cite{W00}.  Finally, we check whether the phase
space volume contraction is equal to the thermodynamic entropy production and
whether the Lyapunov sum rule holds for our mechanism.

\section{Algorithm for the calculation of the Lyapunov exponents} 
\label{lj_alg} 
In a smooth $d$-dimensional system the equations of motion for a phase space
vector ${\bf \Gamma}$, 
\[
\dot{\bf \Gamma}={\bf F}({\bf \Gamma})\; ,
\]
and the corresponding equations of motion for $d$  tangent vectors 
${\mbox{\boldmath $\delta$}\bf \Gamma}=({\mbox{\boldmath $\delta$}\bf r},{\mbox{\boldmath $\delta$}\bf v})$,
\[
\dot{\mbox{\boldmath $\delta$}\bf \Gamma}=\frac{\partial {\bf F}}{\partial {\bf \Gamma}}{\mbox{\boldmath $\delta$}\bf \Gamma} \; ,
\]
are integrated to obtain $d$ Lyapunov exponents 
\begin{equation}
\lambda=\lim_{t\to\infty}\frac{1}{t}\ln\frac{|{\mbox{\boldmath $\delta$}\bf \Gamma}(t)|}
{|{\mbox{\boldmath $\delta$}\bf \Gamma}(0)|} \; .
\label{ljap}
\end{equation}
The Lyapunov exponents are a measure to characterize the stability of the
dynamics \cite{Schu,ER}.  The maximal Lyapunov exponent $\lambda$ measures the
maximal exponential divergence of two initially neighboring points
${\mbox{\boldmath $\delta$}\bf \Gamma}(0)$.  However, during the time evolution every tangent vector
will move into the fastest growing direction due to the instability of the
dynamics.  All these vectors will thus become indistinguishable and their norm
will diverge.  The algorithm of Benettin avoids this problem by a periodic 
Gram-Schmidt reorthonormalization of the tangent vectors thus enabling to
compute the full spectrum of Lyapunov exponents associated to the
$d$-dimensional phase space \cite{BGGS80,WSSV85}.

In the periodic Lorentz gas the time-continuous flow describing the dynamics
of a phase space volume vector ${\bf \Gamma}$ in the bulk is interrupted by a
time-discrete map ${\bf M}$ describing the transformation of ${\bf \Gamma}$ at the moment
of a collision,
\[
{\bf \Gamma'}={\bf M}({\bf \Gamma}) \,.
\]
Dellago and coworkers have developed an algorithm to calculate the Lyapunov
exponents for particle systems with hard sphere interactions and applied it to
the Gaussian thermostated Lorentz gas \cite{DeGP95,DePH96,DePo95}.  Here the
tangent vectors are transformed at the moment of a collision according to the
following rule \cite{DePH96}:
\begin{equation}
{\mbox{\boldmath $\delta$}\bf{ \Gamma'}}=\frac{\partial {\bf M}}{\partial {\bf \Gamma}}{\mbox{\boldmath $\delta$}\bf \Gamma}+
\left[\frac{\partial {\bf M}}{\partial {\bf \Gamma}}{\bf F}({\bf \Gamma})-{\bf F}({\bf M}({\bf \Gamma}))\right]
        \delta\tau_c \,.
\label{ldph}
\end{equation}
Eq.~(\ref{ldph}) is valid for arbitrary systems composed of a flow ${\bf F}$ and a time-discrete map ${\bf M}$.  It takes into account that a trajectory and a
satellite trajectory collide with the disk at different space points and with
a time delay $\delta\tau_c$,
\[
\delta\tau_c=-\frac{({\mbox{\boldmath $\delta$}\bf r},{\bf n})}{({\bf v},{\bf n})}\;,
\]
where ${\bf n}$ is the unit vector perpendicular to the surface at the collision
point. 

\begin{figure}
\epsfxsize=12cm
\centerline{\epsfbox{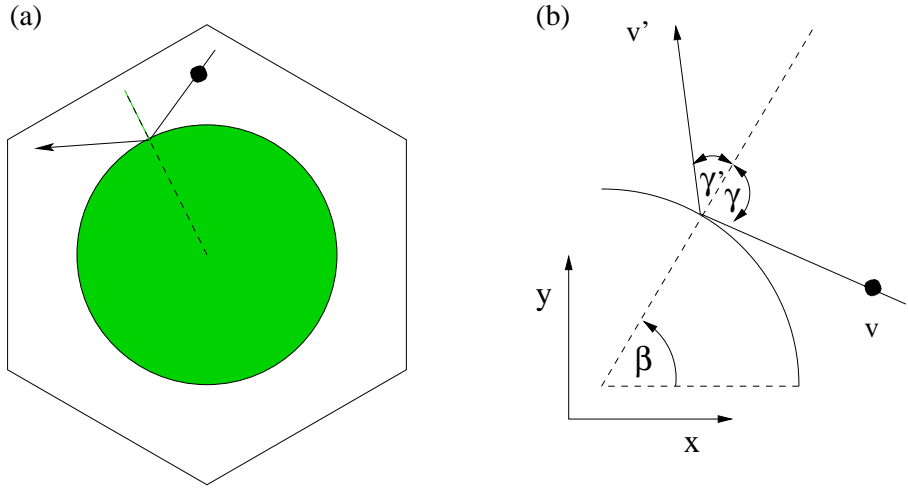}}
\vspace*{0.3cm} 
\caption{(a) Elementary cell of the periodic Lorentz gas on a
triangular lattice. (b) Definition of the relevant variables.}
\label{fig1}
\end{figure}

Before we establish the equations of motions for the tangent vectors of the
Lorentz gas as thermostated by deterministic scattering we briefly summarize
the full equations of motion of Refs.~\cite{KRN98,RKN99} for a particle
described by the phase space vector ${\bf \Gamma}=({\bf r},{\bf v})$. 
In the bulk ${\bf \Gamma}$
evolves according to 
\begin{eqnarray}
{\bf r}&=&\mbox{\boldmath $\varepsilon$}\frac{t^2}{2}+{\bf v}t+{\bf r_0}\nonumber\\
{\bf v}&=&\mbox{\boldmath $\varepsilon$}t+{\bf v_0}
\label{eqm}
\end{eqnarray}
where $\mbox{\boldmath $\varepsilon$}$ is an external electric field of strength
$\varepsilon=|\mbox{\boldmath $\varepsilon$}|$ generating a nonequilibrium
situation.  The basic idea of thermostating by deterministic scattering is now
that at a collision energy is transfered such that the resulting velocity
distribution for the particle is canonical in equilibrium.  In a way, it
results in a deterministic and time-reversible formulation of stochastic
boundary conditions
\cite{RKN99,WKN98}. For this purpose the collision rules 
have been defined as follows: 
The velocity of the particle and its direction of flight are changed
at a collision with the disk according to
\begin{equation}
(\gamma',v')=(X^{-1},Y_\infty^{-1})\circ {\bf B}\circ(X(\gamma),Y_\infty(v))\; ,
\label{coru}
\end{equation}
where $\gamma$ is the angle of incidence, $X(\gamma )=\sin|\gamma|$, ${\bf B}$ is
the baker map \cite{Schu}, and
\begin{equation}
Y_\infty(v) = -\sqrt{\frac{2}{\pi T}}ve^\frac{-v^2}{2T}+\mbox{erf}(\frac{v}{\sqrt{2T}})
\label{iy}
\end{equation} 
with $T$ as a parameter corresponding to the temperature of the particle
$T=<v^2>/2$ at $\varepsilon=0$ in equilibrium.  The geometry of the periodic
Lorentz gas and the relevant variables are shown in Fig.~\ref{fig1}.  To
ensure that the system is time-reversible, the forward baker map ${\bf B}$ acts if
$0\le\gamma\le \pi/2$, and ${\bf B}$ is replaced by its inverse ${\bf B^{-1}}$
if $-\pi/2\le\gamma<0$. To avoid any symmetry breaking induced by this
combination of forward and backward baker map, we alternate their application in
$\gamma$ with respect to the position $\beta$ of the colliding particle on the
circumference. For the spacing between two neighboring disks with the radius
$R=1$ we choose, following the literature
\cite{MH87,DeGP95}, $w\simeq0.2361$. 
Investigating this system in nonequilibrium by switching on an external
electric field $\varepsilon>0$ leads to a nonequilibrium steady state with on
average constant kinetic energy of the particle $<v^2>=const.$, i.~e.~the
system is thermostated.

The equations of motion for the tangent
vectors ${\mbox{\boldmath $\delta$}\bf\Gamma}$ in the bulk can now be derived from Eqs.~(\ref{eqm}) as
\[
\left( \begin{array}{c} {\mbox{\boldmath $\delta$}\bf r'} \\ {\mbox{\boldmath $\delta$}\bf v'}\end{array} \right)= 
\left( \begin{array}{cc} 1 & t \\ 0 & 1 \end{array} \right)
\left( \begin{array}{c} {\mbox{\boldmath $\delta$}\bf r} \\ {\mbox{\boldmath $\delta$}\bf v}\end{array} \right)\;.
\]
The transformation rules for the tangent vectors at the moment of a collision
are obtained by inserting the collision rules for the phase space vector
${\bf \Gamma}$ of Eq.~(\ref{coru}) into Eq.~(\ref{ldph}),
\begin{equation}
\left( \begin{array}{c} {\mbox{\boldmath $\delta$}\bf r'} \\ {\mbox{\boldmath $\delta$}\bf v'}\end{array} \right)= 
\left( \begin{array}{cc}{\bf 1} & {\bf 0} \\ {\bf A} & {\bf B} \end{array} \right)
\left( \begin{array}{c} {\mbox{\boldmath $\delta$}\bf r} \\ {\mbox{\boldmath $\delta$}\bf v}\end{array} \right)+
\left[ \left(\begin{array}{cc}{\bf 1} &{\bf 0} \\ {\bf A} & {\bf B} \end{array} \right)
\left( \begin{array}{c} {\bf v} \\ \mbox{\boldmath $\varepsilon$} \end{array} \right)-
\left( \begin{array}{c} {\bf v'} \\ \mbox{\boldmath $\varepsilon$} \end{array} \right)\right]\delta\tau_c
\label{lrk}\; ,
\end{equation}
where 
\[
A_{ij}=\frac{\partial {v_i}'}{\partial r_j} \hspace{1.cm}\mbox{and}\hspace{1.cm} B_{ij}=\frac{\partial {v_i}'}{\partial v_j}\;.
\]

The components of the submatrix $A$ read

\parbox{7.cm}
{\begin{eqnarray*}
\frac{\partial {v_x}'}{\partial r_x} &=& \frac{r_y {v_y}'}{r^2}[-h_1+1] \\ 
\frac{\partial {v_y}'}{\partial r_x} &=& \frac{r_y {v_x}'}{r^2}[ h_1-1] 
\end{eqnarray*}}
\hfill
\parbox{7.cm}
{\begin{eqnarray*}
\frac{\partial {v_x}'}{\partial r_y} &=& \frac{r_x {v_y}'}{r^2}[ h_1-1]\\ 
\frac{\partial {v_y}'}{\partial r_y} &=& \frac{r_x {v_x}'}{r^2}[-h_1+1] \,
\end{eqnarray*}
}

and the components of the submatrix $B$ are

\parbox{7.cm}
{\begin{eqnarray*}
\frac{\partial {v_x}'}{\partial v_x} &=& \frac{v_y {v_y}'}{v^2} h_1+v_x {v_x}' h_2 \\
\frac{\partial {v_y}'}{\partial v_x} &=& -\frac{v_y {v_x}'}{v^2} h_1+v_x {v_y}' h_2 
\end{eqnarray*}}
\hfill
\parbox{7.cm}
{\begin{eqnarray*}
\frac{\partial {v_x}'}{\partial v_y} &=& -\frac{v_x {v_y}'}{v^2} h_1+v_y {v_x}' h_2 \\
\frac{\partial {v_y}'}{\partial v_y} &=& \frac{v_x {v_x}'}{v^2} h_1+v_y {v_y}' h_2 \,
\end{eqnarray*}}

with

\parbox{7.cm}
{\[
h_1=m_1\frac{v'}{v}\frac{({\bf r},{\bf v})}{({\bf r},{\bf v'})} \nonumber 
\]} 
\hfill
\parbox{7.cm}
{\[
h_2=m_2\frac{v}{v'^3}\rm{e}^{-\frac{v'^2-v^2}{2T}} \, .
\]}

$\{m_1,m_2\}$ are the slopes of the baker map $\{2,0.5\}$, or the slopes of
the inverse baker map $\{0.5,2\}$, respectively.

In the periodic Lorentz gas as thermostated by deterministic scattering the
dynamics of four orthonormal tangent vectors has to be investigated to obtain
four Lyapunov exponents which completely characterize the stability in the
four-dimensional phase space.  Before we present our results for the Lyapunov
spectrum we wish to derive explicit expressions for two other interesting
quantities.

\section{Phase space volume contraction}
\label{spc}
The phase space volume contraction $P$ is equal to the sum of the Lyapunov
exponents \cite{ER,Gasp}, 

\begin{equation}
P=\sum_i \lambda_i \label{pscl}\;.
\end{equation}
In the periodic Lorentz gas as thermostated by deterministic scattering only
the change of a phase space volume element ${\mbox{\boldmath $\delta$}\bf\Gamma}$ at the moment of a
collision as described by Eq.~(\ref{lrk}) contributes to $P$.  The mean
exponential rate of the phase space volume contraction $P$ can then be
calculated according to
\begin{equation}
P=<\ln\left|\frac{\partial{\mbox{\boldmath $\delta$}\bf\Gamma'}}{\partial{\mbox{\boldmath $\delta$}\bf\Gamma}}\right|>
\label{psvc}\;,
\end{equation}
where
\[\left|\frac{\partial{\mbox{\boldmath $\delta$}\bf\Gamma'}}{\partial{\mbox{\boldmath $\delta$}\bf\Gamma}}\right|=\left| \begin{array}{cc} \frac{\partial{\mbox{\boldmath $\delta$}\bf r'}}{\partial{\mbox{\boldmath $\delta$}\bf r}} & {\bf 0} \\ 
\frac{\partial{\mbox{\boldmath $\delta$}\bf v'}}{\partial{\mbox{\boldmath $\delta$}\bf r}} & \frac{\partial{\mbox{\boldmath $\delta$}\bf v'}}{\partial{\mbox{\boldmath $\delta$}\bf v}} \end{array} \right|
=\left(\frac{\partial\delta {r_x}'}{\partial\delta r_x}\frac{\partial\delta {r_y}'}{\partial\delta r_y}-
\frac{\partial\delta {r_x}'}{\partial\delta r_y}\frac{\partial\delta {r_y}'}{\partial\delta r_x}\right)
\left(\frac{\partial\delta {v_x}'}{\partial\delta v_x}\frac{\partial\delta {v_y}'}{\partial\delta v_y}-
 \frac{\partial\delta {v_x}'}{\partial\delta v_y}\frac{\partial\delta {v_y}'}{\partial\delta v_x}\right)\;.
\]
The partial derivatives of Eq.~(\ref{lrk}) read 

\parbox{7.cm}
{\begin{eqnarray*}
\frac{\partial\delta {r_x}'}{\partial\delta r_x} &=& 1+({v_x}'-v_x)\frac{n_x}{({\bf v},{\bf n})} \\ 
\frac{\partial\delta {r_y}'}{\partial\delta r_x} &=& ({v_x}'-v_x)\frac{n_y}{({\bf v},{\bf n})}
\end{eqnarray*}}
\hfill
\parbox{7.cm}
{\begin{eqnarray*}
\frac{\partial\delta {r_x}'}{\partial\delta r_y} &=& ({v_y}'-v_y)\frac{n_x}{({\bf v},{\bf n})} \\ 
\frac{\partial\delta {r_y}'}{\partial\delta r_y} &=& 1+({v_y}'-v_y)\frac{n_y}{({\bf v},{\bf n})} \,
\end{eqnarray*}}

\parbox{7.cm}
{\begin{eqnarray*}
\frac{\partial\delta {v_x}'}{\partial\delta v_x} &=& B_{00}\\
\frac{\partial\delta {v_y}'}{\partial\delta v_x} &=& B_{10}
\end{eqnarray*}}
\hfill
\parbox{7.cm}
{\begin{eqnarray*}
\frac{\partial\delta {v_x}'}{\partial\delta v_y} &=& B_{01}\\
\frac{\partial\delta {v_y}'}{\partial\delta v_y} &=& B_{11}\;.
\end{eqnarray*}}                        

Inserting these expressions into Eq.~(\ref{psvc}) yields
\begin{equation}
P=\frac{<v'^2>-<v^2>}{2T} \;.
\label{pcon}
\end{equation}
The phase space volume contraction depends thus only on the average transfer
of kinetic energy to the reservoir. Equation~(\ref{pcon}) is valid in
equilibrium as well as in nonequilibrium.  An analogous result has been
obtained for collisions with a flat wall in a heat and shear flow thermostated
by deterministic scattering \cite{WKN98}, however, in case of shear the
expression for $P$ turned out to be more complicated.

If $P<0$ the phase space volume typically contracts onto a fractal attractor
in the driven periodic Lorentz gas \cite{MH87,HoMo89,Morr89}.  The geometric
properties of the attractor can be related to the Lyapunov exponents by the
Kaplan-Yorke conjecture, $D_{KY}=D_1$.  Here $D_1$ is the information
dimension \cite{Schu} and $D_{KY}$ is the Kaplan-Yorke dimension defined by
\begin{equation}
D_{KY}=j+\frac{\sum_{i=1}^j \lambda_i}{|\lambda_{j+1}|}\;,
\label{dky}
\end{equation}
where the $\lambda_i$ are ordered by magnitude, $\lambda_1>\lambda_2>\ldots$,
and $j$ is the largest integer for which $\sum_{i=1}^j \lambda_i>0$.

\section{Thermodynamic entropy production and reservoir temperature}
\label{stres}
The macroscopic properties of nonequilibrium steady states can be
characterized by quantities from thermodynamics and statistical physics.  In
this work we want to check whether we can relate the thermodynamic entropy
production $dS$, 
\begin{equation}
dS=\frac{dQ}{T_r}\;,
\label{iep}
\end{equation}
to the phase space volume contraction.  To calculate the thermodynamic entropy
production for thermostating by deterministic scattering we have to calculate
the temperature of the reservoir $T_r$ in nonequilibrium. As discussed in
Ref.~\cite{RKN99}, in nonequilibrium the temperature related to the particle,
or respectively the temperature in the bulk $T_b$ defined via equipartitioning
of energy, is greater than the parametric temperature $T$ in Eq.~(\ref{iy})
and increases with the field strength.  Moreover, $T_b$ is inhomogeneously
distributed in the bulk because the thermostat acts only at the boundary.  In
this subsection we derive an expression for the temperature of the reservoir
$T_r$ similarly to how it has been done in Ref.~\cite{WKN98}.

If we assume equipartitioning of energy of particle and
reservoir at the wall, we can define the reservoir temperature $T_r$
indirectly via the velocity distribution of the particle at the moment of the
collision denoted as $\varrho_{map}$. For sake of simplicity, here we do not
explicitly consider the dependence of $T_r$ on the position $\beta$ of the
colliding particle at the disk. An expression for the temperature of the
reservoir can then be derived from the temperature in the bulk on the basis of
the relation between the map density $\varrho_{map}$ and the time-continuous
density $\varrho$ in the bulk as given by Eq.~(5) in Ref.\ \cite{RKN99},
\begin{equation}
\varrho(v) = const.\frac{\varrho_{map}(v)}{v}\;.
\label{dt}
\end{equation}
The precise derivation of this equation can be found in Sect.\ IIIB2 of Ref.\
\cite{RKN99}. To obtain the expressions
for the velocity fluctuations parallel and perpendicular to the reservoir, the
corresponding equations for the velocity distributions of the normal and
tangential components $v_n$ and $v_t$, respectively, have to be
calculated. For this purpose, first the analogous equation for
$\varrho(\gamma)$ corresponding to Eq.~(\ref{dt}) must be derived. Knowing
that in equilibrium $\varrho(\gamma)=1$ because of symmetry 
and $\varrho_{map}(\gamma)=\cos\gamma$ at the disk 
leads to
\begin{equation}
\varrho(\gamma)=const.\frac{\varrho_{map}(\gamma)}{|\cos(\gamma)|}\;.
\label{gamma}
\end{equation}
Combining these two equations yields the full transformation
\[
\varrho(v)\varrho(\gamma)=\frac{const.}{v|\cos(\gamma)|}\varrho_{map}(v)\varrho_{map}(\gamma)\;.
\]
Changing to local Cartesian coordinates $(v_n,v_t)$ co-rotating with the
position $\beta$ at the disk and applying the transformation
$dv_ndv_t=vdvd\gamma$ results in
\begin{equation}
\varrho(v_n)\varrho(v_t)=\frac{const.}{v|\cos(\gamma)|}\varrho_{map}(v_n)
\varrho_{map}(v_t)
\label{vnvt}
\end{equation}
Noting that $|v_n|=v|\cos(\gamma)|$ and matching the variables on both sides,
Eq.~(\ref{vnvt}) can be decomposed into
\begin{equation}
\varrho(v_t)=\varrho_{map}(v_t) 
\label{tvt}
\end{equation}
with $-\infty<v_t<\infty$ and
\begin{equation}
\varrho(v_n)=\frac{const.}{|v_n|}\varrho_{map}(v_n)
\label{tvn}
\end{equation}
with $0<v_n<\infty$. Before we come to the
reservoir temperature definitions which are based on these densities, we
remark that the disk which serves as the thermal reservoir is fixed and cannot
recognize any current.  In other words, only the kinetic energy of the
particle in the fixed frame of the bulk $2E_{pf}=<v^2>$ is relevant for the
interaction with the reservoir, and no average current needs to be
subtracted. Defining now $[\ldots]$ as the average over $\varrho_{map}$
Eq.~(\ref{tvt}) implies for the tangential component $<v_t^2>=[v_t^2]$ thus
leading to the definition of $T_t$ as
\[ 
T_t=\frac{[v_t^2]+[v_t'^2]}{4}\;.
\] 
Analogously, the average over the map density corresponding to $<v_n^2>$ can
be calculated from Eq.~(\ref{tvn}) to
\[
<v_n^2>=\frac{\int_{-\infty}^\infty v_n^2 \varrho(v_n)dv_n}{\int_{-\infty}^\infty 
\varrho(v_n)dv_n}
=\frac{\int_{-\infty}^\infty |v_n| \varrho_{map}(v_n)dv_n}{\int_{-\infty}^\infty 
\frac{1}{|v_n|} \varrho_{map}(v_n)dv_n}=
\frac{[|v_n|]}{[\frac{1}{|v_n|}]}\;,
\]
where the denominator is obtained from joint normalization over the ingoing
and outgoing fluxes.  $T_n$ is then defined as
\[
T_n=\frac{1}{2}\frac{[|v_n|+v_n']}{[\frac{1}{|v_n|}+\frac{1}{v_n'}]}
\]
with $-\infty<v_n\le 0$ and $0\le v_n'<\infty$. 

The total temperature of the reservoir $T_r$ is consequently the average of
$T_t$ and $T_n$,
\begin{equation}
T_r=\frac{T_t+T_n}{2}\;.
\label{t_res}
\end{equation}
$T_r$ can be calculated as an average over $\beta$ or locally in a small
interval $\Delta\beta$. We note that our result for $T_n$ is slightly
different to the result in Ref.~\cite{WKN98}, which strictly speaking is only
valid if the in- and outgoing densities are symmetrical.

This definition of the temperature is exact in equilibrium, however, in case
of a nonequilibrium situation Eq.~(\ref{dt}) and Eq.~(\ref{gamma}) are not
valid anymore.  A more detailed analysis of these shortcomings leads to the
conclusion that $T_r$ calculated according to Eq.~(\ref{t_res}) will be
greater than the real temperature of the reservoir for higher field strength
\cite{RKun}.  One would only obtain the real temperature of the reservoir if
one would use the correct relation between map density and time-continuous
density in nonequilibrium, and this is not known.  In any case, a lower bound
for the temperature of the reservoir which we denote as $T_{lr}$ can be
calculated by only taking into account the velocity of the particle after a
collision.

\section{Equilibrium} 
The numerical calculation of the Lyapunov spectrum for the Lorentz gas as
thermostated by deterministic scattering according to the method presented in
section \ref{lj_alg} leads to the following result in equilibrium:
$\{\lambda\}=\{1.8695, 0.0104, -0.0104, -1.8695\}$.  These data appear to be
at variance with the fact that in equilibrium two zero Lyapunov exponents have
to exist, one associated with the direction of the flow, and a second one
resulting from the conjugate pairing rule in equilibrium \cite{ER}.  We have
performed the following tests to detect the reason for this discrepancy:
\begin{enumerate}
\item We have numerically calculated the Lyapunov 
exponents by investigating the dynamics of a trajectory and four satellite
trajectories, i.~e., for finite but small distances.  The Lyapunov spectrum
obtained by this method was the same.
\item Changing parameters like the interdisk distance $w$, the parametric
temperature $T$, the dimensionality of the reservoir \cite{KRN98,RKN99}, the
slope of the baker map, and replacing the baker map by more complicated
two-dimensional maps like the cat map or the standard map \cite{Schu} did not
improve the result.
\item We have followed the temporal evolution of two points on the same
trajectory for about 20 collisions without orthonormalization and by choosing
as initial conditions a) that the points are slightly displaced along the
trajectory but have the same velocity, b) that the points have the same
configuration space coordinates but slightly different velocities.  We could
then show that two neutral directions exist corresponding to a) the direction
of the flow and b) to one direction perpendicular to the flow. 
\end{enumerate}

The third test indicates that two zero Lyapunov exponents indeed exist. 
Thus, there must be a numerical problem because of standard Gram-Schmidt 
orthonormalization which
establishes an orthonormal system of the tangent vectors on the basis of the
most unstable direction. To cure that problem, we propose an alternative
method to perform the periodic orthonormalization.  This method establishes an
orthonormal system of the tangent vectors starting from the existing neutral
direction of the flow. Since we are introducing an additional constraint this
way we call it constraint orthonormalization. It consists of the following
steps:

\begin{enumerate}
\item Choose suitable initial conditions for the orthonormal system: The first tangent vector 
is situated in the direction of the flow,
${\mbox{\boldmath $\delta$}\bf\Gamma_1}=(v_{x0}/v_0,v_{y0}/v_0,0,0)$, and the other tangent vectors
are orthonormal to it.
\item At every orthonormalization the first tangent vector is forced to point
in the direction of the flow, ${\mbox{\boldmath $\delta$}\bf\Gamma_1}=(v_x/v,v_y/v,0,0)$. This step
corrects the very small deviations of the first tangent vector from the
direction of the flow resulting from a
collision with the disk, as
will be explained in more detail below.
\item The second, third and fourth tangent vector are orthonormalized again
starting from the first one according to the method of Gram-Schmidt.
\end{enumerate} 

The application of constraint orthonormalization leads to the following
Lyapunov spectrum in equilibrium: $\{\lambda\}=\{1.8695, 0.0000,0.0000,
-1.8695\}$, see also Fig.~\ref{fig3}.  Comparing these results to the
previous ones obtained from the standard method shows that the Gram-Schmidt
orthonormalization led to a wrong result only for the second and third
Lyapunov exponent. The explanation for this numerical problem is as follows:
In equilibrium the average energy transfer to the reservoir is zero. Still, at
any collision energy is transfered either from the particle to the reservoir
or in the opposite direction. According to Eq.~(\ref{pcon}) the phase space
volume thus {\em locally} contracts or expands although the {\em global} phase
space volume contraction is zero. However, as shown by Eq.~(\ref{pscl}) the
phase space contraction is intimately related to the corresponding (un)stable
directions in phase space.  Consequently, the local contractions and
expansions at a collision change the orientation and the norm of the tangent
vectors in a nontrivial way. The Gram-Schmidt orthonormalization reacts to
these changes by turning the corresponding tangent vectors out of the
previously neutral directions. The problem why the Gram-Schmidt procedure does
not converge to the existing two neutral directions at least in the long time
limit could not be completely resolved even by very detailed numerical
investigations of the dynamics. Possibly some kind of resonance phenomenon
between local phase space contraction and expansion at the collision and
Gram-Schmidt orthonormalization after the collision leads to the corresponding
tangent vectors adjusting themselves somewhat symmetrically around these two
neutral directions \cite{failGS}.

In summary, constraint orthonormalization correctly yields a second zero
Lyapunov exponent beside the zero Lyapunov exponent corresponding to the
constrained direction of the flow. In agreement with the on average zero
energy transfer between particle and reservoir the sum of the Lyapunov
exponents and the global phase space volume contraction are zero. Furthermore,
the Lyapunov exponents trivially fulfill the conjugate pairing rule related to
the Hamiltonian character of the dynamics in equilibrium. The probability
density in the Lorentz gas cell is uniform in equilibrium and, as a
consequence, the Kaplan-Yorke dimension is equal to the dimension of the phase
space, $D_{KY}=d=4$.  In addition, the parametric temperature $T$, the
temperature in the bulk $T_b$, and the reservoir temperature $T_r$, are all
equal, $T=T_b=T_r$.

We now turn to an even more detailed analysis of the dynamical instability of
our model system by following ideas summarized in Refs.\ \cite{Gasp,DeHo00}.
If a dynamical system is ergodic, the Lyapunov exponents do not depend on the
initial conditions of the tangent vectors, and thus they only yield
information about the global instability. This implies that Eq.~(\ref{ljap})
provides no direct way to assess the local instability of the system at
specific values of phase space variables like the angle of incidence at a disk
$\gamma$ and the position of the colliding particle $\beta$. In Refs.\
\cite{Gasp,DeHo00}, two slightly different ways have been proposed how to
access information on local instabilities depending on these parameters. Here
we use the approach proposed in Ref.\ \cite{DeHo00} which characterizes the
local deformation of a typical tangent vector $\delta{\bf \Gamma}$ at the moment of
a collision by introducing the quantity
\begin{equation}
\lambda_c(\sin\gamma,\beta)=<\ln\frac{|\delta \bf \Gamma'|}{|\delta \bf \Gamma|}>\;,
\label{lex}
\end{equation}
where the brackets indicate an average over all collisions in a respective
small interval around $\beta$ and/or $\gamma$.  The physical motivation for
defining this quantity is that any tangent vector quickly orients itself into
the direction of fastest growth. Accordingly, the full memory about the
maximum instability of the system is contained in the orientation of the
tangent vector thus representing a ``needle'' in phase space which very
sensitively measures the local changes of the stability at a
collision. Therefore, this quantity is a very sensitive function of $\gamma$
and $\beta$. In Ref.\ \cite{DeHo00} this quantity has been called a local
Lyapunov exponent, however, this term has also been used in the literature to
indicate the dependence of the Lyapunov exponents Eq.~(\ref{ljap}) on initial
conditions in case the dynamics is non-ergodic \cite{ER}. To avoid possible
confusion, and by following Ref.\ \cite{Gasp} where very related quantities
have been defined, here we denote $\lambda_c$ as the local stretching rates of
the system. Note that the clever and very simple definition by Eq.~(\ref{lex})
makes at least the maximum local stretching rate directly accessible to
computer simulations. In contrast, in Ref.\ \cite{Gasp} the full spectrum of
these rates has been defined in a proper co-moving coordinate system. This
makes their definition more convenient in mathematical terms, but also less
accessible for straightforward numerical computations. Both these different
definitions are related via coordinate transformations \cite{PGun}.
Unfortunately, local stretching rates are not coordinate-invariant thus
yielding different values depending on their precise definition, even in
conjugate dynamical systems.

\begin{figure}[htbp]
\epsfxsize=16cm
\centerline{\epsfbox{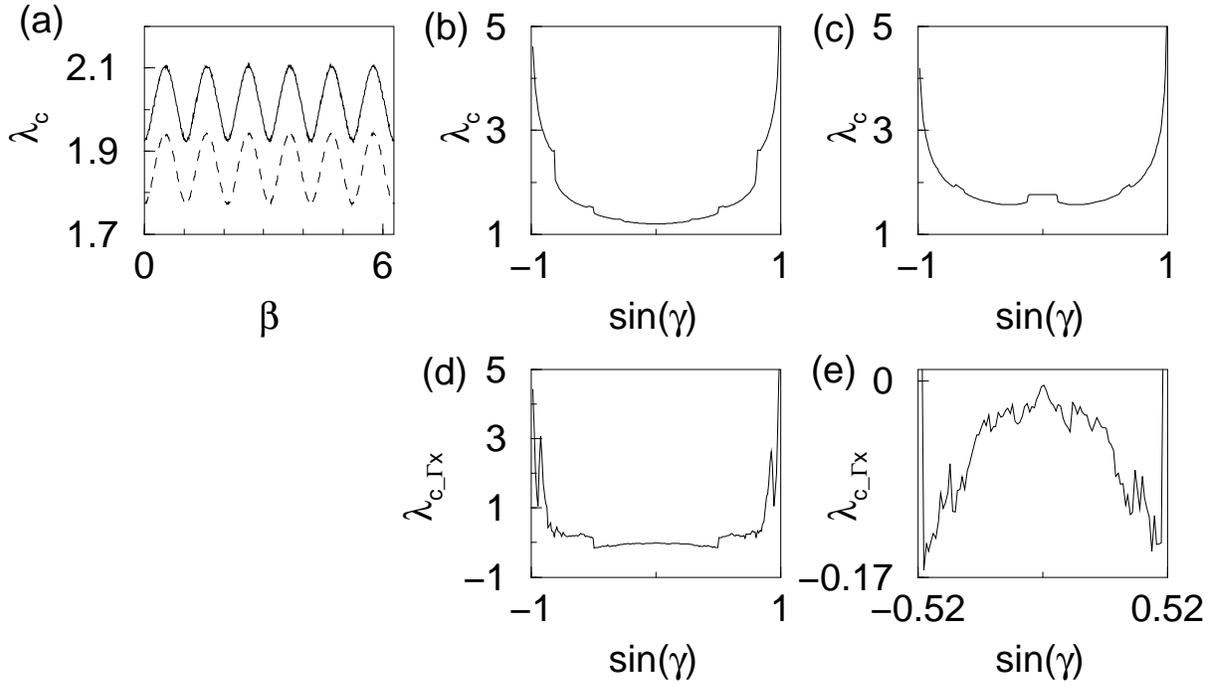}}
\caption{Local stretching rate in equilibrium: (a) 
thermostating by deterministic scattering (solid curve), elastic collisions
(dashed curve) for all $\gamma$; (b)-(e) thermostating by deterministic
scattering, (b) $|\beta|<0.00001$, (c) $|\beta-\pi/2|<0.00001$; (d),(e)
refined local stretching rate, (d) $|\beta-\pi/2|<0.00001$, (e) enlarged
sector of (d).}
\label{fig2}
\end{figure}

$\lambda_c$ as a function of $\beta$ for thermostating by deterministic
scattering in comparison to elastic collisions is presented in
Fig.~\ref{fig2}(a) showing that the conventional hard disk Lorentz gas and our
thermostated version of it share the same properties. The maxima/minima of
$\lambda_c(\beta)$ correspond to the directions of maximal/minimal distances
between neighboring disks, respectively.  Results for the conventional Lorentz
gas in a more detailed view of phase space, i.e., $\lambda_c(\sin(\gamma))$
for $|\beta|<0.00001$ as presented in Ref.~\cite{DeHo00}, have shown that the
local stretching rate $\lambda_c$ is a singular function of $\sin(\gamma)$
(see also Ref.\ \cite{Gasp}).  Related results for $\lambda_c(\sin(\gamma))$
for the Lorentz gas as thermostated by deterministic scattering are presented
in Fig.~\ref{fig2}. The curve in Fig.~\ref{fig2}(b) looks qualitatively very
similar to the curve in Fig.~1 of \cite{DeHo00}.  However, the numerical
results for $\lambda_c(\sin(\gamma))$ of our system are not sufficiently
accurate \cite{CPU} to study the existing discontinuities on a finer scale, as
it has been done in Fig.~2 of Ref.\ \cite{DeHo00}.  To perform such
investigations in a slightly more detailed way we looked at the refined,
decomposed local stretching rate $\lambda_{c_{\Gamma_x}}=<\ln(|\delta
\Gamma_x'|)/(|\delta \Gamma_x|)>$ which characterizes the deformation of the
x-component of a tangent vector only.  The results for
$\lambda_{c_{\Gamma_x}}$ are presented in Figs.~\ref{fig2} (d) and (e).
Fig.~\ref{fig2}(e) shows an enlarged sector of (d) where one can see a roughly
symmetric profile composed of maxima and minima on a fine scale.  The apparent
symmetry of most of these peaks suggests that these oscillations are not due
to numerical errors. We consider this as an indication that for the Lorentz
gas as thermostated by deterministic scattering at least the refined local
stretching rate $\lambda_{c_{\Gamma_x}}$ could be a singular function of
$\sin(\gamma)$. It may be somewhat surprising that such specific dynamical
properties of the conventional, unthermostated Lorentz gas persist in our
thermostated system as well. However, this leads to the conclusion that the
geometric instability of the system is more important for these
characteristics than the one resulting from the modifications related to our
specific scattering mechanism.

\section{Nonequilibrium}
\label{lne}
\begin{figure}[htbp]
\epsfxsize=16cm
\centerline{\epsfbox{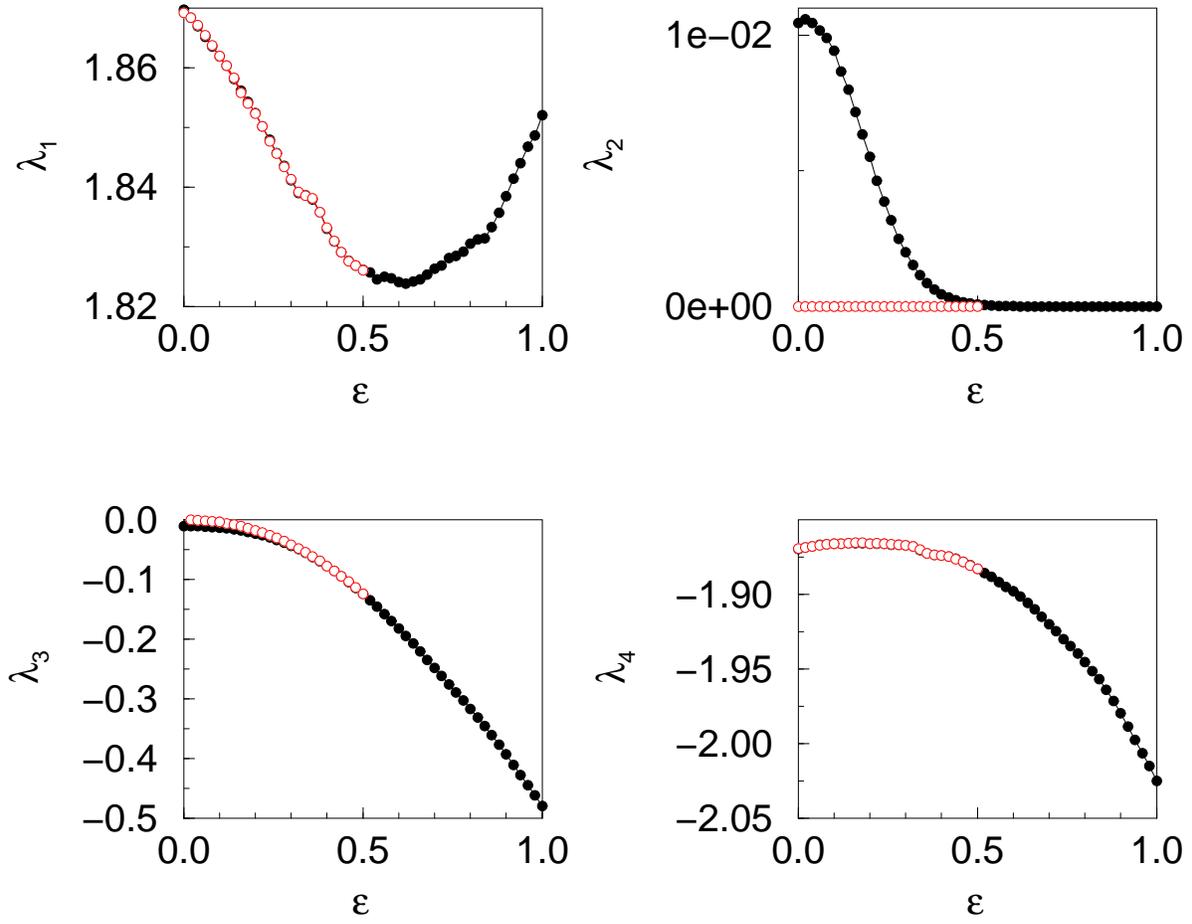}}
\caption{Field dependence of the four Lyapunov exponents in nonequilibrium; 
filled circles: Gram-Schmidt orthonormalization, empty circles: constraint
orthogonalization as defined in the text. In this and in the following figures
the error for the results is less then $5\cdot10^{-4}$.}
\label{fig3}
\end{figure}

In nonequilibrium we choose the electric field such that $\varepsilon_x>0,
\varepsilon_y=0$. The field accelerates the particle, and energy is transfered
on average to the disk resulting in a nonequilibrium steady state. As a
consequence, the global phase space volume contraction given by
Eq.~(\ref{pcon}) is negative.  The detailed dependence of the Lyapunov
spectrum on the field strength is shown in Fig.~\ref{fig3}, where both
results from the standard method as well as from the constraint method are
presented. Only one zero Lyapunov exponent exists in nonequilibrium associated
with the direction of the flow. For higher field strength unconstraint
Gram-Schmidt orthonormalization correctly turns the second tangent vector in
the direction of the flow, $\lambda_2<10^{-4}$ for $\varepsilon_x>0.5$. 
 
To obtain the correct Lyapunov exponents for $\varepsilon_x<0.5$ in
nonequilibrium, we apply a suitably adjusted version of constraint
orthonormalization as used in equilibrium: In order to achieve that two points
on a trajectory stay on the same trajectory after a collision, their initial
states and velocities have to be chosen such that the points have the same
velocity at the moment of the collision. This condition leads to the
components for the first tangent vector
${\mbox{\boldmath $\delta$}\bf \Gamma_1}=\{v_{x0},v_{y0},\varepsilon_x,0\}$.  
Note that ${\mbox{\boldmath $\delta$}\bf\Gamma}$
is not normalized here. The other steps are then the same as in equilibrium.

The Lyapunov spectrum as a function of the field strength obtained from
constraint orthonormalization is also presented in Fig.~\ref{fig3}.  As in
equilibrium, the results of the two methods differ only for the second and
third Lyapunov exponent for $\varepsilon_x<0.5$.  The differences for the
third Lyapunov exponent are of the same size as the differences for the second
Lyapunov exponent.  The second Lyapunov exponent obtained by the new method is
zero for all field strengths corresponding to the constrained tangent vector
in the direction of the flow.  The third Lyapunov exponent decreases with
increasing field strength which is related to the dominant energy transfer in
the direction from the particle to the disk.  The dependence of the third and
of the fourth Lyapunov exponent on the field strength appears to be a power
law, which is a behavior that has also been observed for the Gaussian
thermostated Lorentz gas for small enough field strength \cite{MDI96}.
According to Pesins theorem
\cite{ER}, the only positive Lyapunov exponent is equal to the Kolmogorov-Sinai
entropy $h_{KS}$,
\[
h_{KS}=\sum_i \lambda_i^+ \;.
\]
Interestingly, its curve is nonmonotonous.  For small field strength, the
dynamics in configuration space appears to be dominated by the fact that the
trajectory of the particle is getting adjusted in the direction of the field,
and the Kolmogorov-Sinai entropy decreases.  For higher field strength the
increasingly disordered dynamics in velocity space related to an increase of
the bulk temperature $T_b$ seems to become more important, and the
Kolmogorov-Sinai entropy increases.  The same field dependence of the
Kolmogorov-Sinai entropy has been observed in a shear flow as thermostated by
deterministic scattering \cite{W00}. It would be interesting to know whether
this is a general property appearing in field-driven system as thermostated by
deterministic scattering. In contrast to this observation, the
Kolmogorov-Sinai entropy monotonically decreases for the Gaussian thermostated
Lorentz gas because the constraint of the bulk thermostat onto the dynamics
increases with increasing field strength
\cite{DeGP95,DettmannMorrissRondoni95}.
Whether the irregularities on the fine scale in Fig.~\ref{fig3} are a
property of the dynamics or whether they are numerical fluctuations could not
be decided on the basis of the present data.

\begin{figure}[htbp]
\epsfxsize=16cm
\centerline{\epsfbox{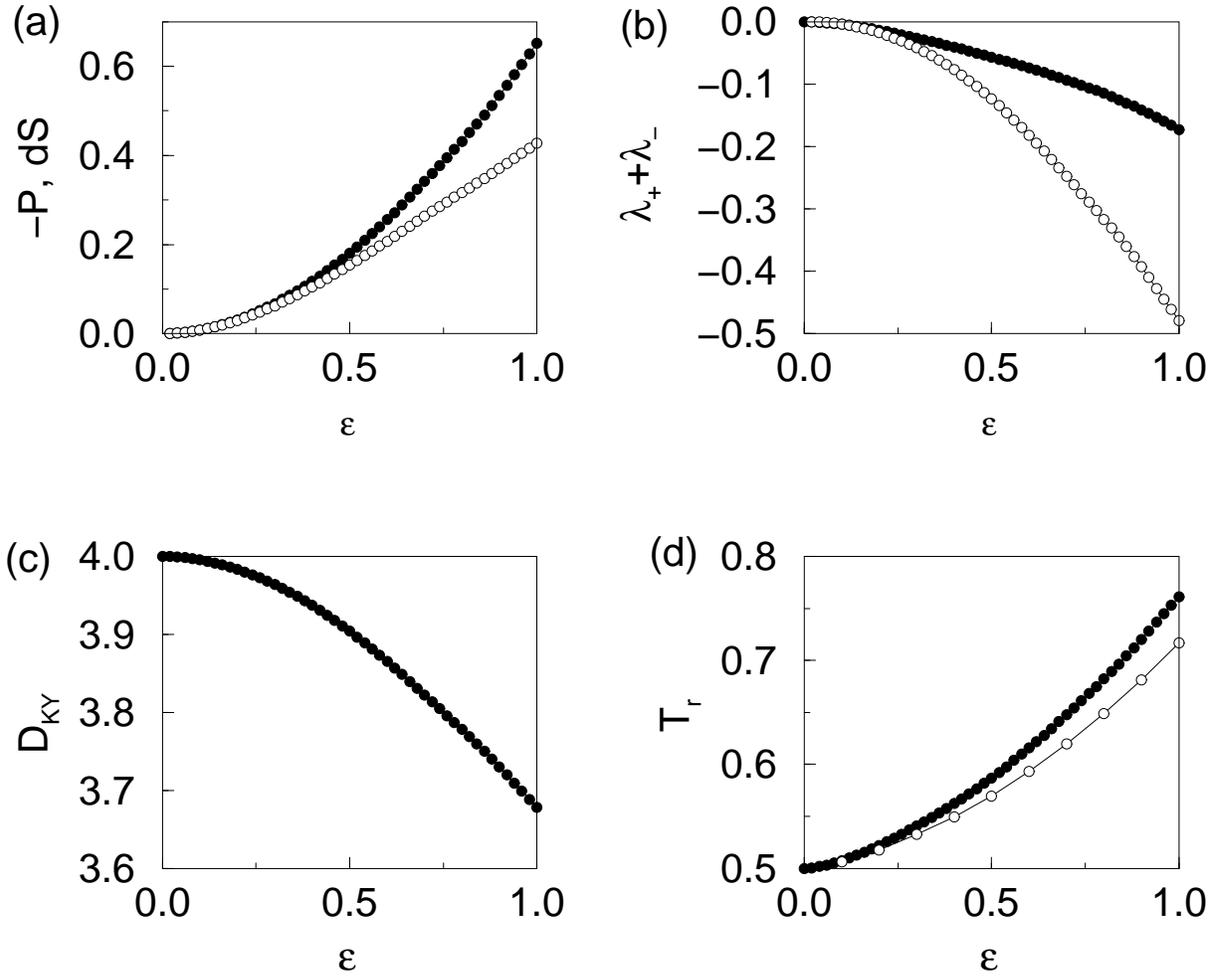}}
\caption{Field dependence of various important quantities in nonequilibrium:
(a) black: phase space contraction rate $-P$, white: irreversible entropy
production $dS$; (b) conjugate pairs of Lyapunov
exponents, black: $\lambda_1+\lambda_4$, white: $\lambda_2+\lambda_3$; (c)
Kaplan-Yorke dimension $D_{KY}$; (d) reservoir temperature $T_r$ (black) and
its lower bound $T_{lr}$ (white).}
\label{fig4}
\end{figure}

The sum of the Lyapunov exponents is negative and according to
Eq.~(\ref{pcon}) equal to the phase space volume contraction $P$. As presented
in Fig.~\ref{fig4}(a), $P$ decreases with increasing field strength.  The
density of the attractor remains phase space filling but shows a nonuniform
and complicated structure as shown in the Poincar\'e section in Fig.~4(a) of
Ref.~\cite{KRN98}.  Therefore we can assume that the Hausdorff dimension $D_0$
is equal to the dimension of the phase space, $D_0=d=4$, as is also the case
for Gaussian thermostated periodic Lorentz gases.  In contrast, the
Kaplan-Yorke dimension $D_{KY}$ defined by Eq.~(\ref{dky}) is not an integer
anymore, as presented in Fig.~\ref{fig4}(c).  This provides quantitative
evidence for the fractal structure of the attractor according to the
conjecture $D_{KY}=D_1$.

Some of the conventional deterministic and time-reversible bulk thermostats
fulfill the conjugate pairing rule saying that the Lyapunov exponents can be
grouped into pairs such that $\lambda_++\lambda_-=const.$ \cite{MoDe98,ECM}.
Fig.~\ref{fig4}(b) shows that the conjugate pairing rule does not hold for our
model. However, this does not come as a big surprise because it is well-known
that even conventional thermostats do not exhibit conjugate pairing if
thermostated at the boundaries \cite{PH89,De00}.  

\begin{figure}[htbp]
\epsfxsize=16cm
\centerline{\epsfbox{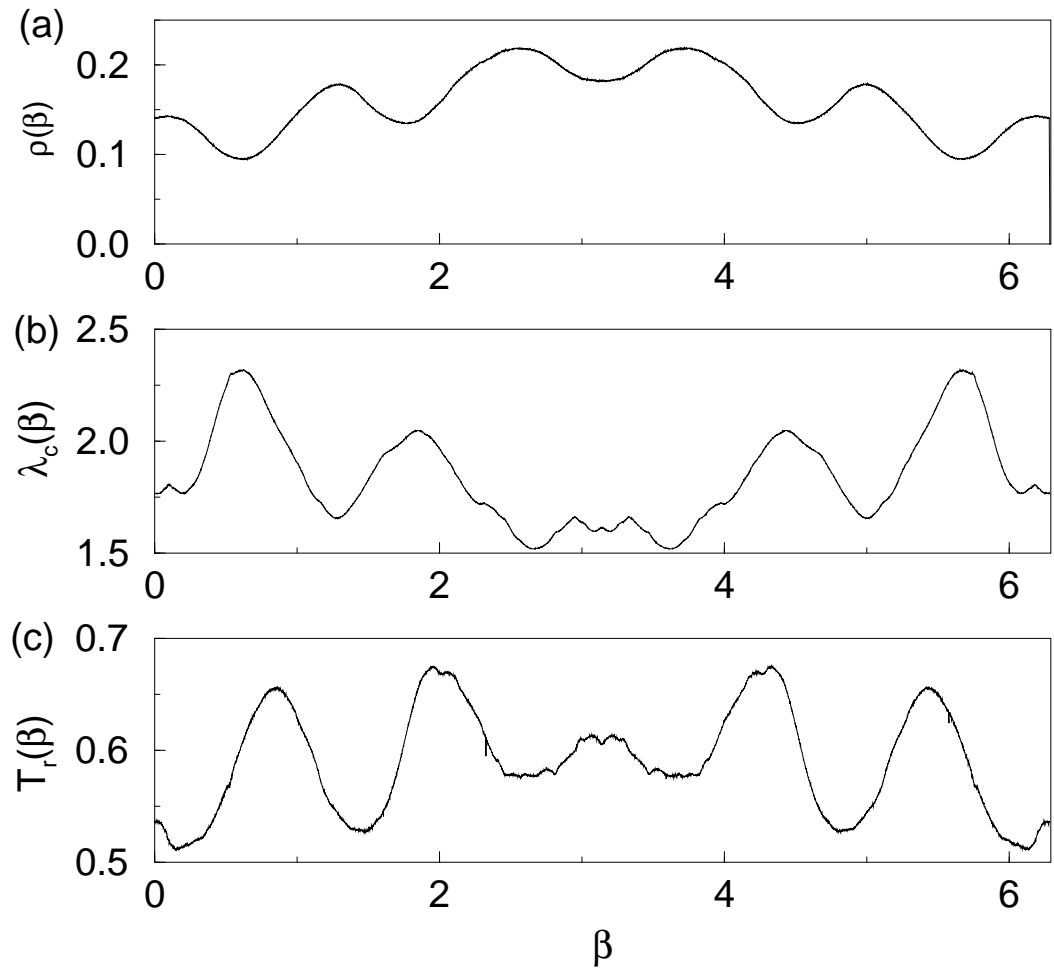}}
\caption{Different quantities evaluated at the collision of a particle with
the disk at field strength $\varepsilon_x=0.5$: (a) density at a collision
$\varrho(\beta)$; (b) local stretching rate $\lambda_c(\beta)$, (c) reservoir
temperature $T_r(\beta)$.}
\label{fig5}
\end{figure}

The local expansion rate $\lambda_c(\beta)$ as defined in Eq.~(\ref{lex}) is
presented in Fig.~\ref{fig5}(b) and can be compared with $\varrho(\beta)$
shown in Fig.~\ref{fig5}(a).  One still recovers remnants of the periodic
equilibrium distribution of $\lambda_c$, see Fig.~\ref{fig2}(a).  However,
they are strongly deformed by the anisotropy induced by the field, and the
maxima and minima are much more pronounced.  In contrast to equilibrium, there
exist two absolute maxima, one around $\beta\approx\pi/6$ and one around
$\beta\approx2\pi-\pi/6$, and an absolute minimum around $\beta\approx\pi$.
The maxima and minima of $\lambda_c(\beta)$ occur just opposite to the maxima
and minima of $\varrho(\beta)$. This is in agreement with the physical
interpretation that a more unstable dynamics leads to a more dilute particle
density in phase space. To extend the comparison, the temperature of the
reservoir $T_r$ as a function of $\beta$ calculated according to
Eq.~(\ref{t_res}) is presented in Fig.~\ref{fig5}(c). The distribution of
peaks in $\lambda_{c}(\beta)$ and $T_r(\beta)$ is very similar.  This might be
related to the fact that both quantities illustrate somewhat irregular
behavior: $\lambda_c$ characterizes the instability of the dynamics and $T_r$
is equal to the mean kinetic energy of the degrees of freedom of the
reservoir.  The analogous $\beta$-dependence of $\lambda_{c}$ and $T_r$ points
again to a close relation between dynamical system theory and statistical
mechanics. At $\beta\approx\pi$ both the distributions of $\lambda_c$ and
$T_{r}$ show a more complicated structure.  This is probably a consequence of
the dynamics being directed parallel to the field resulting in the global
minimum of $\lambda_c$ on a coarse scale, whereas for other $\beta$ the
dynamics is more chaotic. For more detailed views of the phase space in terms
of the local stretching rate, in analogy to Fig.~\ref{fig2} in equilibrium, we
could not get qualitative good results \cite{CPU}.  Thus, whether
$\lambda_c(\beta)$ is a singular function in nonequilibrium on a fine scale
remains an open question.

The dependence of $T_r$ on the field strength according to the definition in
Eq.~(\ref{t_res}), in which $T_r$ is averaged over $\beta$, and the lower
bound $T_{lr}$ as defined below this equation are shown in Fig.~\ref{fig4}(d).
In particular, the results for $T_{lr}(\varepsilon)$ confirm that
$T_r$ is always greater in nonequilibrium than the parametric temperature $T$,
$T_r>T$.

\begin{figure}[htbp]
\epsfxsize=16cm
\centerline{\epsfbox{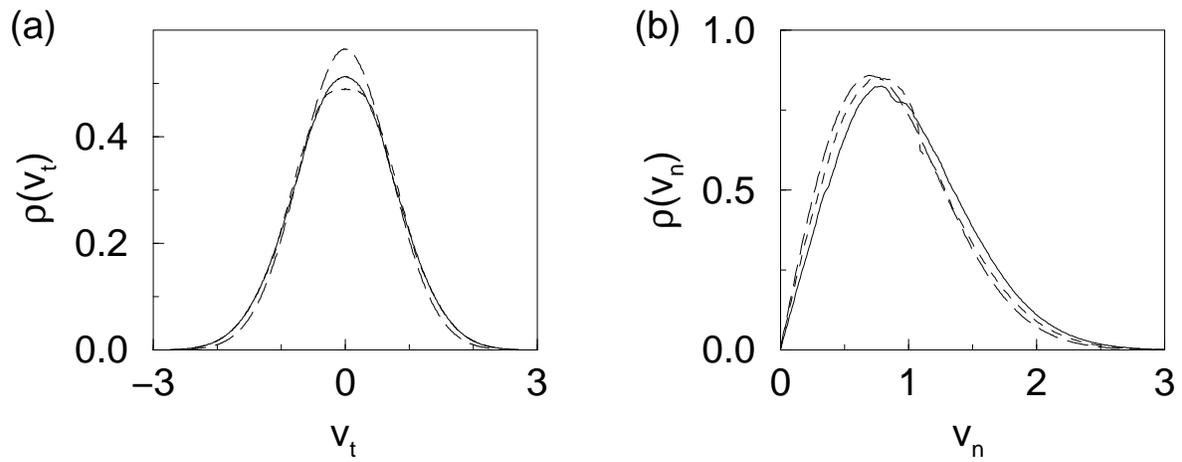}}
\caption{Probability density at the moment of a collision for (a) tangential
velocity $v_t$ and (b) normal component $v_n$ at field strength
$\varepsilon_x=0.5$; solid curve: density before collision, dashed curve:
density after collision, long dashed curve: density in equilibrium.}
\label{fig6}
\end{figure}

More detailed information related to the deviations between reservoir
temperature and parametric temperature are obtained by studying the map
densities $\varrho_{map}(v_t)$ and $\varrho_{map}(v_n)$ at a collision as
represented in Fig.~\ref{fig6}. The deviations between ingoing and outgoing
densities in both cases are reminiscent of an average transfer of kinetic
energy from particle to reservoir, as it is necessary to compensate the influx
of energy caused by the electric field to obtain a nonequilibrium steady
state. However, in case of the periodic Lorentz gas taking the thermodynamic
limit leaves the system precisely as it is. Consequently, there is no
thermodynamic way to get rid of the difference between ingoing and outgoing
velocity distribution. But these differences are the dynamical reason why in
nonequilibrium the reservoir temperature $T_r$ is typically not equal to the
parametric temperature $T$, because this would only be the case if both
distributions would be converging to the (local) equilibrium distribution in
the thermodynamic limit, as included in these figures. This aspect will become
important for understanding our results on the relation between phase space
contraction and entropy production below.

\begin{figure}[htbp]
\epsfxsize=16cm
\centerline{\epsfbox{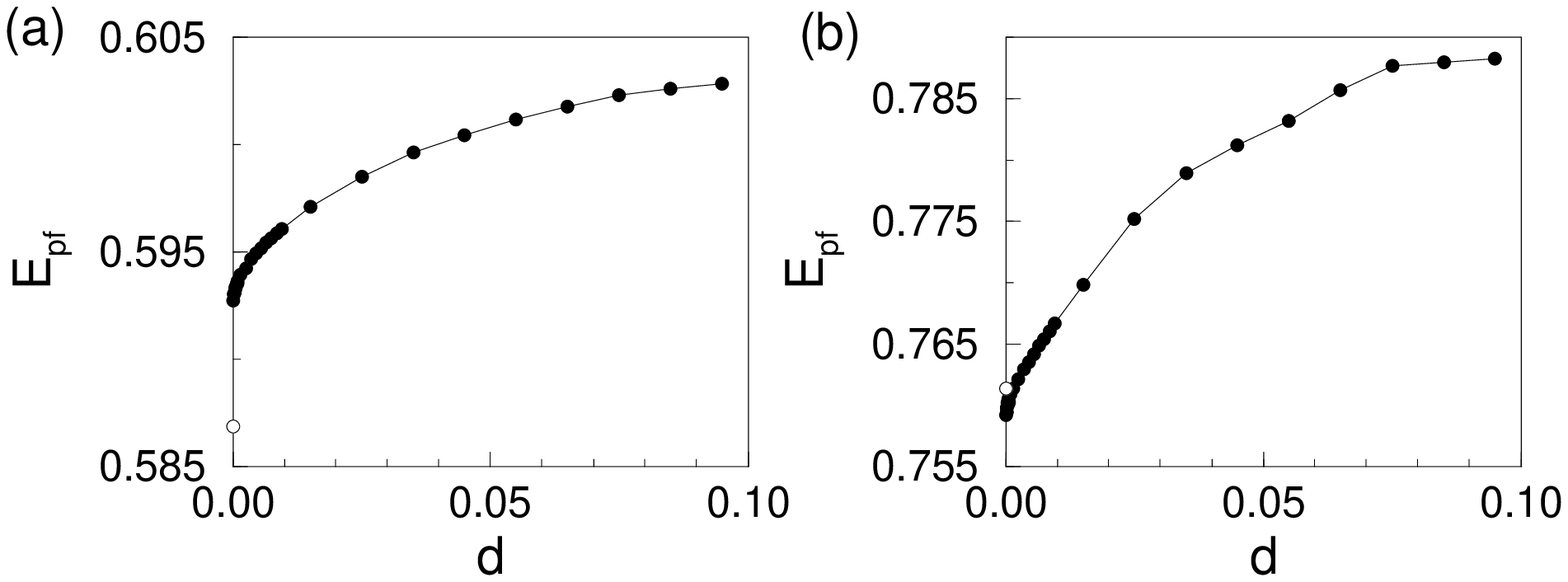}}
\caption{Profile of the full kinetic energy $E_{pf}$ of the particle in the bulk, black
$E_{pf}$, white reservoir temperature $T_r$ (a) field strength
$\varepsilon_x=0.5$, (b) $\varepsilon_x=1$.}
\label{fig7}
\end{figure}

In Fig.~\ref{fig7}, the kinetic energy of the particle in the bulk averaged
over $\beta$, $E_{pf}=<v^2>/2$, is presented as a function of the distance $d$
from the disk. The profile of $E_{pf}$ is inhomogeneous as expected.  For
$d\to0$, $E_{pf}$ should approach the temperature of the reservoir defined via
equipartitioning of energy thus providing an alternative definition of the
reservoir temperature based on the bulk dynamics in the inner ring around the
disk. However, it is not possible to safely extrapolate to this limiting value
on the basis of the present data \cite{CPU}.  Comparing $E_{pf}$ for $d\to 0$
with $T_r$ shows in particular that $T_r>E_{pf}$ for $\varepsilon_x=1$
excluding convergence via extrapolation, i.e., the assumptions made in the
derivation of Eq.~(\ref{t_res}) do not hold in nonequilibrium, but at least
they yield a reasonable estimate. Note that the kinetic energy in this limit
is still always greater than the lower bound for the reservoir temperature,
$E_{pf}>T_{lr}$, which will be important for our following discussion of
entropy production.

The external driving force $\mbox{\boldmath $\varepsilon$}$ performs work
on the system and causes a macroscopic flow characterized by a positive
conductivity $\sigma>0$
\cite{RKN99}. At the same time, work is transformed into heat 
and in turn removed by the thermostat leading to a positive thermodynamic
entropy production. Starting from Eq.~(\ref{iep}), the irreversible entropy
production in the bulk $dS$ is easily computed by defining the heat
production $dQ$ as the change of the kinetic energy of the particle in
the bulk, $d E_{pf}/dt$, and feeding in the bulk equations of motion
Eq.~(\ref{eqm}). This leads to the well-known expression of entropy production
via Joule heating
\begin{equation}
dS=\frac{\varepsilon_x<v_x>}{T_r}\;.
\label{jou}
\end{equation}
The numerical result for the field dependence of the thermodynamic entropy
production according to this equation is presented in Fig.~\ref{fig4}(a). 

On the other hand, as discussed above the heat produced in the bulk must leave
as an outward flux across the walls absorbed by the thermal reservoir.
Correspondingly, computing the average change of the kinetic energy during a
free flight from the equations of motion yields
\begin{equation}
\varepsilon_x<v_x>=\frac{<v^2>-<v'^2>}{2} \; ,
\end{equation}
where the right hand side is just the average transfer of kinetic energy at a
collision.  Inserting this result into Eq.~(\ref{jou}) leads to
\begin{equation}
dS=\frac{<v^2>-<v'^2>}{2T_r}\;.
\label{ieptds}
\end{equation}
Comparing now Eq.~(\ref{ieptds}) with Eq.~(\ref{pcon}) yields the important
result that the identity between thermodynamic entropy production and
phase space volume contraction does not hold for the Lorentz gas as
thermostated by deterministic scattering.  Instead, these two quantities just
differ by the factor $T/T_r(\varepsilon_x)$,
\begin{equation}
dS=-P\frac{T}{T_r(\varepsilon_x)} \; .
\label{eq}
\end{equation}
To explicitly compare these two quantities the field dependence of $-P$ is
also presented in Fig.~\ref{fig4}(a). As we have discussed above, there is
some ambiguity in defining the reservoir temperature $T_r$, however, we
emphasize that all our applied definitions and bounds lead to the result that
$-P$ and $dS$ are inherently different in nonequilibrium. This is also clear
from the fact how the thermostat works in our model, as explained above.

As has been done in conventional thermostats, starting from Eq.~(\ref{eq}) a
relation between the electrical conductivity and the phase space volume
contraction can now be established by using Eq.~(\ref{jou}) and replacing the
average current according to the definition of the conductivity
\[
\sigma=\frac{<v_x>}{\varepsilon_x}
\]
yielding
\[
\sigma=\frac{-TP}{\varepsilon^2_x}=\frac{-T\sum\lambda}{\varepsilon^2_x}\;.
\]
This equation is formally identical to the Lyapunov sum rule obtained for the
conventional thermostats.  The only difference is the constant factor $T$
which, for conventional thermostats, corresponds to the temperature of the
reservoir. If the Lyapunov sum rule applies, it shows that macroscopic
transport can be directly understood in terms of the microscopic dynamics
characterized by the sum of the Lyapunov exponents.

However, we remark that the existence of a Lyapunov sum rule in thermostated
systems being of the simple type as one above rather seems to be the exception
than the rule: For example, a difference between phase space volume
contraction and thermodynamic entropy production has also been obtained for a
shear flow as thermostated by deterministic scattering \cite{WKN98}.  For this
system the expressions for $P$ and $dS$ can be rather complicated.
Consequently, the Lyapunov sum rule does not hold, and a similar relation has
not been found in addition.  Furthermore, already a variation of the
Nos\'e-Hoover and of the Gaussian thermostat did not lead to an identity
between $P$ and $dS$ implying the invalidity of the Lyapunov sum rule as well,
as discussed in Ref.~\cite{RKH00,RKun}.

In general, the relation between phase space volume contraction and
thermodynamic entropy production, and the corresponding relation between
transport coefficient and Lyapunov exponents, will depend on the details of
the microscopic energy transfer between particle and reservoir. 
Based on our studies in Refs.\ \cite{KRN98,RKH00,RKN99,WKN98}, we
conclude that an identity between $P$ and $dS$ appears only to be valid for
what might be called ``ideal'' thermostats meaning that energy is exchanged
between subsystem and reservoir by sufficiently simple coupling rules as they
are provided, for example by conventional Gaussian and Nose-Hoover
thermostats.

\section{Conclusions}
In this work we have numerically calculated the Lyapunov exponents for the
Lorentz gas thermostated by deterministic scattering.  The Gram-Schmidt
orthonormalization, a fundamental ingredience of the standard method to
calculate Lyapunov exponents, led to a wrong result for the Lyapunov spectrum
by applying this thermostat.  We modified this method by imposing an
additional constraint, summarized as constraint orthonormalization, and found
results which are in agreement with expectations from dynamical systems
theory. We wish to remark that the phenomenon causing our numerical
difficulties is reminiscent of an inelastic collision of a particle with a
hard disk, as it is also modeled in granular materials by using restitution
coefficients. Thus, applying constraint orthonormalization might be helpful
for exactly computing Lyapunov spectra in low-dimensional systems of granular
type as well. On the basis of the Lyapunov exponents further quantities have
been calculated to characterize the nonequilibrium steady state.  The
comparison of the results obtained for thermostating by deterministic
scattering with the ones known for conventional thermostats leads to the
following conclusions:

1. The sum of the Lyapunov exponents for thermostating by deterministic
scattering is negative in nonequilibrium in agreement with the phase space
volume contraction onto an attractor.  For thermostating by deterministic
scattering only one Lyapunov exponent is zero in nonequilibrium related to the
direction of the flow. Similar results could be expected for the Nos\'e-Hoover
thermostated Lorentz gas where the calculation of the Lyapunov exponents have
not yet been performed. In contrast, for the Gaussian thermostated Lorentz gas
two Lyapunov exponents are zero in nonequilibrium because the thermostat keeps
the kinetic energy of the particle strictly constant.

2. The Kaplan-Yorke dimension calculated on the basis of the Lyapunov exponent
is not an integer in nonequilibrium providing quantitative evidence that the
attractor of thermostating by deterministic scattering in the periodic Lorentz
gas exhibits a fractal structure analogous to the conventional bulk
thermostats.

3. The identity between thermodynamic entropy production and phase space
volume contraction does not hold for thermostating by deterministic
scattering. Instead, these two quantities differ by a field dependent factor.
The reason for this difference is that the temperature of the reservoir of
thermostating by deterministic scattering depends on the field strength, in
contrast to Gaussian and Nos\'e-Hoover thermostats. This result is important,
since this identity was accepted up to now as a general characterization of
nonequilibrium steady states generated by deterministic and time-reversible
thermostats.

4. Surprisingly, although there is no identity we could still establish a
relation between conductivity and Lyapunov exponents for thermostating by
deterministic scattering.  This equation is formally identical to the Lyapunov
sum rule for conventional thermostats. As far as we know, our model thus
provides a first example of a system where there is no identity, but where
nevertheless there is a simple relation between transport coefficients and
dynamical instabilities similar to conventional thermostats.

In summary, we find that the existence of fractal attractors in nonequilibrium
steady states are common features which thermostating by deterministic
scattering shares with conventional thermostats. Physically speaking, the
fractal character reflects the extreme rarity of nonequilibrium states
relative to equilibrium ones. To look for additional common properties of all
deterministic and time-reversible thermostats remains an important question,
which is intimately related to obtaining a general characterization of
nonequilibrium steady states.  Such a characterization might result in a more
general relation between quantities of thermodynamic interest and the
indicators of dynamical chaos at the microscopic level, from which the
relations obtained for the thermostating mechanisms considered above could
appear as special cases.

{\bf Acknowledgments:} We are indebted to C.~Dellago for his assistance
concerning the numerical calculation of the Lyapunov exponents, and we thank
Prof.\ G.~Nicolis for his ongoing support and encouragement in this research.
Thanks go to J.~Wiersig for helpful proof-reading. K.R.\ thanks the European
Commission for a TMR grant under contract no.~ERBFMBICT96-1193 and the
foundation ``D.~\&~A. van Buuren'' for financial support. R.K.\ wishes to
acknowledge support from the MPIPKS for this work in form of a distinguished
postdoctoral fellowship.

\begin{thebibliography}{10}
\bibitem{failGS}
This failure of the Gram-Schmidt orthonormalization was not detected in related
  calculations for a shear flow as thermostated by deterministic scattering
  \cite{W00}. We suspect that, because of the many elastic particle-particle
  collisions in the bulk, in this system it has quantitatively negligible
  consequences.
\bibitem{CPU}
One week CPU time was not sufficient to increase the quality of the results.
\end{thebibliography}
\end{document}